Length of inflation and WMAP data in the case of power-law inflation

Shiro Hirai* and Tomoyuki Takami**

Department of Digital Games, Faculty of Information Science and Arts,
Osaka Electro-Communication University
1130-70 Kiyotaki, Shijonawate, Osaka 575-0063, Japan
*Email: hirai@isc.osakac.ac.jp
**Email: takami@isc.osakac.ac.jp

**Abstract**

The effect of the length of inflation on the power spectra of scalar and tensor perturbations is estimated using the power-law inflation model with a scale factor of $a(\eta) = (-\eta)^p = t^q$. Considering various pre-inflation models with radiation-dominated or matter-dominated periods before inflation in combination with two matching conditions, the power spectrum of curvature perturbations at large scales is calculated. Comparison of the derived angular power spectrum with the Wilkinson Microwave Anisotropy Probe (WMAP) data reveals a possibility that the WMAP can be explained by the finite length of inflation model if the length of inflation is near 60 *e*-folds and $q \geq 200$.

PACS number: 98.80.Cq

## 1. Introduction

Inflation is an important concept in cosmology and is strongly supported by recent satellite-based experiments. However, consistent and natural models of inflation from the point of view of particle physics have yet to be established. Recently released data obtained by the Wilkinson Microwave Anisotropy Probe (WMAP) has allowed a number of cosmological parameters to be fixed precisely. Although the WMAP data can be almost satisfactorily explained by the ΛCDM model [1], there remains some inconsistency in the suppression of the spectrum at large angular scales ($\ell = 2$), as well as the running of the spectral index and oscillatory behavior in the spectrum. The effect of this small discrepancy on the inflation models and pre-inflation physics is thus an interesting problem.

Many attempts have been made to explain the gap between the WMAP data and the ΛCDM model. As simple slow-roll inflation models appear to be unable to adequately explain the three features mentioned above [2], double inflation models and more complicated models have been considered [3]. The suppression of the spectrum seen in the WMAP data at large angular scales has been addressed by many models, including a finite-sized universe model with nontrivial topology or a closed universe [4], models involving new physics [5], models involving a cutoff [6], and models with an initially kinetics-dominated region or some other region [7]. The study of the trans-Planckian problem [8] is very important in this regard. In this study, however, an alternative approach is considered based on the assumption of a finite inflation of close to 60 *e*-folds in length. In this case, a state of pre-inflation must be considered. This article discusses the effect of such a model with a finite length of inflation and pre-inflation on the power spectrum and angular power spectrum.

Our group recently considered the effect of the initial condition in inflation on the power spectrum of curvature perturbations [9]. Based on the physical conditions before inflation, the possibility exists that the initial state of scalar perturbations in inflation is not simply the Bunch-Davies state, but a more general squeezed state [10]. A formula for the power spectrum of curvature perturbations having any initial condition in inflation was derived for this model as a familiar formulation multiplied by a factor indicating the contribution of the initial condition.

The previous papers [11,12] considered finite inflation models in which inflation began at a certain time, preceded by pre-inflation as a radiation-dominated or scalar matter-dominated period. The matching conditions for the scalar perturbations are essential in the cases of inflation [13] and an ekpyrotic scenario [14]. Calculations of the power spectrum were made for two matching conditions; one in which the gauge potential and its first $\eta$-derivative are continuous at the transition point, and one in which the transition occurs on a hyper-surface of constant energy as proposed by Deruelle and Mukhanov [13]. The differences among the models and matching conditions were investigated in detail by calculating the power spectrum of curvature perturbations, the spectral index, and the running spectral index. This analysis revealed that when the length of inflation is finite, the power spectra can be expressed by decreasing functions at super large scales, and that in cases of radiation-dominated and matter-dominated pre-inflation with the special matching condition, the power spectra oscillate from large to small scales.

In this paper, in order to investigate the effect of such properties of the power spectrum on the angular power spectrum, the angular power spectrum is calculated for the same pre-inflation models. That is, the model assumes inflation of finite length in the case of power-law inflation using a scale factor of $a(\eta) = (-\eta)^p = t^q$ with either radiation-dominated or scalar matter-dominated pre-inflation and the two matching conditions. These toy pre-inflation models are used for the following reasons. If the universe is very hot in pre-inflation, even if a massive particle dominates the energy density of the universe, the universe can be approximated as being dominated by radiation-like matter. Alternatively, it is natural that the scalar field that causes inflation dominates the energy density of the universe. The power spectra of curvature perturbations and gravitational waves can be calculated analytically for the two models and matching conditions. Using the CMBFAST code [15] and the derived the power spectra, the angular power spectra are calculated for all cases over various lengths of inflation and values of $p$. Comparison of the derived angular power spectra with the WMAP data yields some interesting results.

This paper is organized as follows. In section 2, formulae for the power spectra of curvature perturbations and gravitational waves are re-derived for any initial condition in inflation. In section 3, the pre-inflation models are discussed, and in section 4, the power spectra for curvature perturbations and gravitational waves are re-derived for the two models and two matching conditions. In section 5, using the derived formula, the angular power spectra are calculated and compared with WMAP data. In section 6, the results obtained in the present study are discussed at length.

## 2. Scalar and tenser perturbations

The formula for the power spectrum of curvature perturbations in inflation is derived here for any initial condition by applying a commonly used method [16]. This formula was originally derived in Refs. [17,9], but as it represents a critical result, it is derived again here for completeness. As a background spectrum, we consider a spatially flat Friedman-Robertson-Walker (FRW) universe described by metric perturbations. The line element for the background and perturbations is generally expressed as [18]

$$\mathrm{d}s^2 = a^2(\eta)\{(1+2A)\mathrm{d}\eta^2 - 2\partial_i B\mathrm{d}x^i\mathrm{d}\eta - [(1-2\Psi)\delta_{ij} + 2\partial_i\partial_j E + h_{ij}]\mathrm{d}x^i\mathrm{d}x^j\}, \quad (1)$$

where $\eta$ is the conformal time. The functions $A$, $B$, $\Psi$ and $E$ represent the scalar perturbations, and $h_{ij}$ represent tensor perturbations. The density perturbation in terms of the intrinsic curvature perturbation of comoving hypersurfaces is given by

$$\Re = -\Psi - \frac{H}{\dot{\phi}}\delta\phi \,, \quad (2)$$

where $\phi$ is the inflaton field, $\delta\phi$ is the fluctuation of the inflaton field, $H$ is the Hubble expansion parameter, and $\Re$ is the curvature perturbation. Overdots represent derivatives with respect to time $t$, and the prime represents the derivative with respect to the conformal time $\eta$. Introducing the gauge-invariant potential $u \equiv a(\eta)(\delta\phi + (\dot{\phi}/H)\Psi)$ allows the action for scalar perturbations to be written as [19]

$$S = \frac{1}{2}\int \mathrm{d}\eta\mathrm{d}^3x\,\{(\frac{\partial u}{\partial\eta})^2 - c_s^2\,(\nabla u)^2 + \frac{Z''}{Z}u^2\}, \quad (3)$$

where $c_s$ is the velocity of sound, $Z = a\dot{\phi}/H$, and $u = -Z\Re$. The field $u(\eta,\boldsymbol{x})$ is expressed using annihilation and creation operators as follows.

$$u(\eta,\boldsymbol{x}) = \frac{1}{(2\pi)^{3/2}} \int d^3k\{\ \boldsymbol{a}_{\boldsymbol{k}}\, u_k(\eta) + \boldsymbol{a}_{-\boldsymbol{k}}{}^{\dagger} u^*{}_k(\eta)\}\, e^{-\boldsymbol{kx}}. \tag{4}$$

The field equation for $u_k$ is derived as

$$\frac{\mathrm{d}^2 u_k}{\mathrm{d}\eta^2} + (c_s^2\, k^2 - \frac{1}{Z}\frac{\mathrm{d}^2 Z}{\mathrm{d}\eta^2})\ u_k = 0. \tag{5}$$

The solution to $u_k$ satisfies the normalization condition $u_k \mathrm{d}u_k^* / \mathrm{d}\eta - u_k^* \mathrm{d}u_k / \mathrm{d}\eta = i$. We consider the power-law inflation $a(\eta) \approx (-\eta)^p$ $(= t^{\,p/(p+1)})$, in which case equation (5) is written as

$$\frac{\mathrm{d}^2 u_k}{\mathrm{d}\eta^2} + (k^2 - \frac{p(p-1)}{\eta^2}) u_k = 0, \tag{6}$$

where $c_s^2 = 1$ in the scalar field case. The solution for equation (6) is written as

$$f_k^I(\eta) = i\frac{\sqrt{\pi}}{2} e^{-ip\pi/2} (-\eta)^{1/2} H^{(1)}_{-p+1/2}(-k\eta), \tag{7}$$

where $H^{(1)}_{-p+1/2}$ is the Hankel function of the first kind with order $-p + 1/2$. As a general initial condition, the mode function $u_k(\eta)$ is assumed to be

$$u_k(\eta) = c_1 f_k^I(\eta) + c_2 f_k^{I*}(\eta), \tag{8}$$

where the coefficients $c_1$ and $c_2$ obey the relation $|c_1|^2 - |c_2|^2 = 1$. The important point here is that the coefficients $c_1$ and $c_2$ do not change during inflation. In ordinary cases, the field $u_k(\eta)$ is considered to be in the Bunch-Davies state, i.e., $c_1 = 1$ and $c_2 = 0$, because as $\eta \to -\infty$, the field $u_k(\eta)$ must approach plane waves, e.g., $e^{-ik\eta}/\sqrt{2k}$.

Next, the power spectrum is defined as follows [16].

$$<\Re_k(\eta), \Re_l^*(\eta)> = \frac{2\pi^2}{k^3} P_\Re \, \delta^3(\boldsymbol{k}-\boldsymbol{l}), \tag{9}$$

where $\Re_k(\eta)$ is the Fourier series of the curvature perturbation $\Re$. The power spectrum $P_\Re^{1/2}$ is then written as follows [16].

$$P_\Re^{1/2} = \sqrt{\frac{k^3}{2\pi^2}} \left|\frac{u_k}{Z}\right|. \tag{10}$$

Using the approximation of the Hankel function, the power spectrum of the leading and next-leading corrections of $-k\eta$ in the case of squeezed initial states can be written as

$$P_\Re^{1/2} = (2^{-p}\,(-p)^p \frac{\Gamma(-p+1/2)}{\Gamma(3/2)} \frac{1}{m_P^2} \frac{H^2}{|H'|})|_{k=aH} \;(1-\frac{(-k\eta)^2}{2(1+2p)})|\, c_1\, e^{-ip\pi/2} + c_2\, e^{ip\pi/2}\;|$$

$$= (\frac{H^2}{2\pi\dot{\phi}})\,|_{k=aH}\; |\, c_1\, e^{-ip\pi/2} + c_2\, e^{ip\pi/2}\;|. \tag{11}$$

where $\Gamma(-p+1/2)$ represents the Gamma function. This formula differs slightly from Hwang's formula [17] due to the introduction of the term $e^{-ip\pi/2}$ into equation (7), as required such that in the limit $\eta \to -\infty$, the field $u_k(\eta)$ must approach plane waves. The quantity $C(k)$ is defined as

$$C(k) = c_1\, e^{-ip\pi/2} + c_2\, e^{ip\pi/2}. \tag{12}$$

If $|C(k)| = 1$, the leading term of $-k\eta$ in equation (11) can be written as $P_\Re^{1/2} = (\frac{H^2}{2\pi\dot{\phi}})\,|_{k=aH}$ [20]. However, equation (11) implies that if, under some physical circumstances, the Bunch-Davies state is not adopted as the initial condition of the field $u_k$, the possibility exists that $|C(k)| \neq 1$. In section 4, the value of $|C(k)|$ is calculated using a number of pre-inflation models.

Next, tensor perturbations are considered. In equation (1), $h_{ij}$ represent gravitational waves. Under the transverse traceless gauge, the action of gravitational waves in the linear approximation is given by

$$S = \frac{1}{2} \int \mathrm{d}^4 x \, \{ (\frac{\partial h}{\partial \eta})^2 - (\nabla h)^2 + \frac{a''}{a} h^2 \}, \tag{13}$$

where $h$ is the transverse traceless part of the deviation of $h_{ij}$ and represents the two independent polarization states of the wave ($h_+ , h_\times$). As is well known, the action of gravitational waves in the linear approximation is in the same form as that for a real massless, minimally coupled scalar field [21]. The field $h(\eta,\boldsymbol{x})$, can thus be expanded in terms of the annihilation and creation operators $\boldsymbol{a_k}$ and $\boldsymbol{a_{-k}}^\dagger$, i.e.,

$$h(\eta,\boldsymbol{x}) = \frac{1}{(2\pi)^{3/2} a} \int d^3 k \, \{ v_{\mathrm{k}}(\eta) \, \boldsymbol{a_k} + v_k^*(\eta) \, \boldsymbol{a_{-k}}^\dagger \} \, e^{-i \boldsymbol{kx}}, \tag{14}$$

where $v_k(\eta)$ is the solution of

$$\frac{\mathrm{d}^2 v_k}{\mathrm{d}\eta^2} + (k^2 - \frac{1}{a} \frac{\mathrm{d}^2 a}{\mathrm{d}\eta^2}) v_k = 0. \tag{15}$$

For inflation, the power-law inflation given by $a(\eta) \approx (-\eta)^p$ ($= t^{p/(p+1)}$) is considered. The solution to equation (15) is the same as that for equation (7). It is assumed that as a general initial condition, the mode function $v_k(\eta)$ is written as

$$v_{\mathrm{k}}(\eta) = c_{g1} \, f_k^I(\eta) + c_{g2} \, f_k^{I*}(\eta), \tag{16}$$

where the coefficients $c_{g1}$ and $c_{g2}$ obey $|c_{g1}|^2 - |c_{g2}|^2 = 1$. Similar to the case for scalar perturbations, equation (16) describes a squeezed state.

The power spectrum under this condition is defined as

$$< v_k(\eta)\, v_l^*(\eta) >= \frac{\pi m_\mathrm{P}^2 a^2}{16k^3}\, P_\mathrm{g}\, \delta^3(\boldsymbol{k}\text{-}\boldsymbol{l}), \tag{17}$$

where $m_\mathrm{P}$ is the Planck mass. The power spectrum for gravitational waves $P_\mathrm{g}^{1/2}$ is then written as [16]

$$P_\mathrm{g}^{1/2} = \frac{4\sqrt{k^3}}{m_\mathrm{P}\sqrt{\pi}\, a} |v_k|. \tag{18}$$

The power spectrum of the leading and next-leading corrections of $|-k\eta|$ in the case of squeezed initial states can be written as

$$P_g^{1/2} = \Big(\frac{2^{-p+1}}{\sqrt{\pi}}(-p)^p \frac{\Gamma(-p+1/2)}{\Gamma(3/2)} \frac{H}{m_P}\Big)\Big|_{k=aH} \Big(1\text{-}\frac{(-k\eta)^2}{2(1+2p)}\Big) \times |c_{g1}\, e^{-ip\pi/2} + c_{g2}\, e^{ip\pi/2}|, \tag{19}$$

where $P_g^{1/2}$ is multiplied by a factor $\sqrt{2}$ for the two polarization states. The quantity $C_g(k)$ is then defined as

$$C_g(k) = c_{g1}\, e^{-ip\pi/2} + c_{g2}\, e^{ip\pi/2}. \tag{20}$$

This formula differs slightly from Hwang's formula [17] for the same reason as in the scalar case. The power spectrum for gravitational waves is obtained by multiplying the familiar formation (i.e., $c_{g1} = 1$ and $c_{g2} = 0$) by the quantities $|C_g(k)|$, similar to the formula derived in the case of scalar perturbations.

**3. Pre-inflationary cosmological models**

The effect of the length of inflation is examined using simplified models of pre-inflation as an illustration. Here, the pre-inflation model is considered to consist simply of a radiation-dominated period or a scalar matter-dominated period. A simple cosmological model is assumed, as defined by

Pre-inflation: $a_{\rm P}(\eta)=b_1(-\eta-\eta_j)^r$,

Inflation: $a_{\rm I}(\eta)=b_2(-\eta)^p$, (21a)

where

$$\eta_j=(\frac{r}{p}-1)\eta_2, \quad b_1=(\frac{p}{r})^r(-\eta_2)^{p-r}b_2. \qquad (21b)$$

The scale factor $a_{\rm I}(\eta)$ represents ordinary inflation. If $p<-1$, the inflation is power-law inflation ($p=-10/9$, $a(t)=t^{10}$), and if $p=-1$, the inflation is de-Sitter inflation, which is not considered here. Inflation is assumed to begin at $\eta=\eta_2$ and end at $\eta=\eta_3$. In pre-inflation, for the case $r=1$, the scale factor $a_{\rm p}(\eta)$ indicates that pre-inflation is a radiation-dominated period, whereas for the case $r=2$, the scale factor $a_{\rm p}(\eta)$ indicates a scalar matter-dominated period. Here, the period of inflation is assumed to be sufficiently long, that is, in the plot of $|C(k)|^2$, and $|C_g(k)|^2$, the start of inflation $\eta_2$ is chosen as the time at which perturbations of the current Hubble horizon size exceed the Hubble radius in inflation.

**4. Calculation of power spectrum**

Using the pre-inflation models and taking account of the matching conditions, the quantities $|C(k)|$ and $|C_g(k)|$ (the power spectra) are calculated as follows.

### 4.1 Radiation-dominated period before inflation

In the case of a radiation-dominated period before inflation, the scale factor becomes $a_\mathrm{p} = b_1(-\eta - \eta_j)$, i.e., $r = 1$. A difference between the scalar and tensor perturbations (gravitational waves) occurs in the radiation-dominated period, that is, the equations for the fields become different. In the case of gravitational waves, the solution to equation (15) is written as

$$f_k^R(\eta) = \frac{1}{\sqrt{2k}} e^{-ik(\eta+\eta_i)} . \tag{22}$$

On the other hand, in the case of curvature perturbations, the field equation $u_k$ can be written as equation (5) in the radiation-dominated period. In this case, $Z = a_\mathrm{p}[2(\mathcal{H}^2 - \mathcal{H}')/3]^{1/2}/(c_\mathrm{s}\mathcal{H})$ [19,22], where $\mathcal{H} = a_\mathrm{p}'/a_\mathrm{p}$. Fixing the value of $c_s^2$ at 1/3, the solution of equation (5) becomes

$$f_k^{SR}(\eta) = \frac{3^{1/4}}{\sqrt{2k}} e^{-ik(\eta+\eta_j)/\sqrt{3}} . \tag{23}$$

For simplicity, it is assumed that the mode functions of the radiation-dominated period can be written as equation (22) for gravitational waves, and as equation (23) for scalar perturbations. In power-law inflation, the equations can be written as equation (6). The general mode functions in inflation can then be written as

$$u_k^I(\eta) = c_1\ f_k^I(\eta) + c_2\ f_k^{I*}(\eta). \tag{24}$$

$$v_k^I(\eta) = c_{g1}\ f_k^I(\eta) + c_{g2}\ f_k^{I*}(\eta), \tag{25}$$

To fix the coefficients $c_1$, $c_2$, $c_{g1}$ and $c_{g2}$, we use the matching condition in which the mode function and first η-derivative of the mode function are continuous at the transition time $\eta = \eta_2$ ($\eta_2$ is the beginning of inflation). The coefficients $c_1$, $c_2$, $c_{g1}$ and $c_{g2}$ can then be calculated as

$$c_1 = \frac{\sqrt{\pi}}{23^{3/4}\sqrt{2z}}\ e^{i(p\pi/2+z/\sqrt{3}p)}((3\text{-}3p\text{-}i\sqrt{3}\,z)\,H^{(2)}_{-p+1/2}(z)\ \text{-}\ 3z\,H^{(2)}_{-p+3/2}(z)), \tag{26}$$

$$c_2 = \frac{\sqrt{\pi}}{23^{3/4}\sqrt{2z}}\ e^{i(-p\pi/2+z/\sqrt{3}p)}((3\text{-}3p\text{-}i\sqrt{3}\,z)\,H^{(1)}_{-p+1/2}(z)\ \text{-}3z\ \ H^{(1)}_{-p+3/2}(\mathrm{z})), \tag{27}$$

$$c_{g1} = -\frac{\sqrt{\pi}}{2\sqrt{2z}}\ e^{i(p\pi/2+z/p)}((\text{-}1\text{+}p\text{+}i\mathrm{z})\,H^{(2)}_{-p+1/2}(z) + z\,H^{(2)}_{-p+3/2}(z)), \tag{28}$$

$$c_{g2} = -\frac{\sqrt{\pi}}{2\sqrt{2z}}\ e^{i(-p\pi/2+z/p)}((\text{-}1\text{+}p\text{+}iz)\,H^{(1)}_{-p+1/2}(z) + z\ \ H^{(1)}_{-p+3/2}(\mathrm{z})), \tag{29}$$

where $z = -k\eta_2$. The quantities $C(k)$ and $C_g(k)$ can then be derived from equations (12) and (20) as follows.

$$C(k) = -\frac{\sqrt{\pi}}{23^{3/4}\sqrt{2z}}\ e^{iz/\sqrt{3}p}$$

$$\{(\text{-}3\text{+}3p\text{+}\mathrm{i}\sqrt{3}\,\mathrm{z})(H^{(1)}_{-p+1/2}(z)\text{+}H^{(2)}_{-p+1/2}(z))\text{+}3z(H^{(1)}_{-p+3/2}(z)\text{+}H^{(2)}_{-p+3/2}(z))\}, \tag{30}$$

$$C_g(k) = -\frac{\sqrt{\pi}}{2\sqrt{2z}}\ e^{iz/p}$$

$$\{(-1+p+iz)(H^{(1)}_{-p+1/2}(z)+H^{(2)}_{-p+1/2}(z))+z(H^{(1)}_{-p+3/2}(z)+H^{(2)}_{-p+3/2}(z))\}. \tag{31}$$

In the case of $z \to 0$, $|C(k)|^2$ and $|C_g(k)|^2$ become zero (i.e., $z^{-2p}$) in both cases. In the case of $z \to \infty$ (from large to small scales), the quantities $|C(k)|^2$ and $|C_g(k)|^2$ are approximately given by

$$|C(k)|^2 \cong \frac{1}{\sqrt{3}}(2+\cos(p\pi+2z)), \tag{32}$$

$$|C_g(k)|^2 \cong 1-\frac{p(p-1)\cos(p\pi+2z)}{2z^2}. \tag{33}$$

The behavior differs between the scalar and tensor cases. $|C(k)|^2$ oscillates around $2/\sqrt{3} \cong 1.1547$, but the amplitude does not depend on $p$ of the leading order. Numerically, $1/\sqrt{3} \le |C(k)|^2 \le \sqrt{3}$ , but $|C_g(k)|^2$ becomes 1. The quantities $|C(k)|^2$ and $|C_g(k)|^2$ are plotted as a function of $z$ $(=-k\eta_2)$ in Figs. 1 and 2 for the case $p = -500/499$.

In these figures, the length of inflation is assumed to be 60 *e*-folds. Perturbations of the current Hubble horizon size for other lengths of inflation are also shown. For example, if the length of inflation is $a$ times longer, the value of $z \approx a$ gives the perturbations of the current Hubble horizon size. In the case of the curvature perturbations, if a longer inflation exists, the quantity $|C(k)|^2$ behaves according to equation (32) from the super-horizon scales to small scales, that is, it oscillates around $2/\sqrt{3}$ (Fig. 1). It is important to note that even a very long inflationary period cannot remove the vibration of $|C(k)|^2$ around $2/\sqrt{3}$ .

Next, consider the ratio of gravitational waves to the curvature perturbations on the power spectrum. The ratio $R(k)$ is written as

$$R(k) = \frac{P_g}{P_{\Re}} = \frac{16(p+1)}{p} \frac{|C_g(k)|^2}{|C(k)|^2} . \tag{34}$$

The first term $16(p+1)/p$ represents the contribution of power-law inflation. In the case of $z \to 0$ (super-large scales), $R_c(k) \approx 1/\sqrt{3}$ ($R_c = |C_g(k)|^2 / |C(k)|^2$), and in the case of $|z|>>1$, $R_c(k)$ becomes $\sqrt{3}/(2+\cos(p\pi + 2z))$.

### 4.2 Scalar matter-dominated period before inflation

When the period before inflation is dominated by scalar matter (given by the inflaton field $\phi$), the scale factor becomes $a_p = b_1(-\eta - \eta_j)^2$, i.e., $r = 2$. In this case, equations. (5) and (15) become the same, and the quantities $C(k)$ and $C_g(k)$ are of the same form. By a similar procedure as used in section 4.1, the quantities $C(k)$ and $C_g(k)$ can be derived as follows.

$$C(k) = C_g(k) = \frac{-i\sqrt{\pi}}{8\sqrt{2z^3}} e^{2iz/p} \{(p^2 + 4z(i+z) - 2p(1+iz))(H^{(1)}_{-p+1/2}(z) + H^{(2)}_{-p+1/2}(z))$$

$$+ 2(p-2iz)z(H^{(1)}_{-p+3/2}(z) + H^{(2)}_{-p+3/2}(z))\}. \tag{35}$$

The quantities $|C(k)|^2$ are proportional to $z^{-2-2p}$ when $z \to 0$ ($k \to 0$), and so $|C(k)|$ becomes zero. For $z \to \infty$ (for large to small scales), $|C(k)|^2$ and $|C_g(k)|^2$ are obtained as

$$|C(k)|^2 = |C_g(k)|^2 \cong 1 - \frac{p(p-2)\cos(p\pi + 2z)}{4z^2} . \tag{36}$$

In this case, $|C(k)|^2 = |C_g(k)|^2 \cong 1$, which differs from the case for the radiation-dominated period before inflation.

To compare the case of the radiation-dominated period with the case of the scalar matter-dominated period, $|C(k)|^2$ $(=|C_g(k)|^2)$ as given above is plotted as a function of $z$ in figure 3 for the case $p = -500/499$. The form undergoes very little change with $p$. As the quantities $C_g(k)$ and $C(k)$ are the same, the ratio $R_c = |C_g(k)|^2 / |C(k)|^2 = 1$, and the ratio $R(k) = 16(p+1)/p$.

### 4.3. Matching conditions

One of two matching conditions for scalar perturbations was used in sections 4.1 and 4.2 above. The first is a matching condition in which the gauge potential and its first $\eta$ -derivative are continuous at the transition point (Condition A). This matching condition allows the initial condition of pre-inflation to be decided rationally, that is, in the limit $\eta \to -\infty$, the field $u_k(\eta)$ approaches plane waves. The second matching condition is that of Deruelle and Mukhanov [13] for cosmological perturbations, which requires that the transition occurs on a hyper-surface of constant energy (Condition B).

The quantity $|C(k)|$ is calculated here for Deruelle's matching condition, which dictates that $\Phi$ and $\xi$ ($\Re$) are continuous at the transition time $\eta_2$, and the difference between the two matching conditions is investigated. The parameters $\Phi$ and $\xi$ can be written as [13,19]

$$\Phi = (-\mathscr{H}^2 + \mathscr{H}')\,(\frac{u}{Z})'/(c_s^2 k^2 \mathscr{H}), \tag{37}$$

$$\zeta = (\mathscr{H}\Phi' - \Phi\,\mathscr{H}' + 2\mathscr{H}^2\,\Phi)/(\mathscr{H}^2 - \mathscr{H}'), \tag{38}$$

where $Z$ is written as $a(\eta)\,(\mathscr{H}^2 - \mathscr{H}')^{1/2}/(c_s\sqrt{4\pi G}\,\mathscr{H})$. As $\Phi$ and $\xi$ can be written in terms of $u_k(\eta)$, the coefficients $c_1$ and $c_2$ are obtained as follows. In the period of pre-inflation, the mode function $u_k(\eta)$ is derived from equation (5), and $\Phi_P$ and $\xi_P$ can be obtained using the relations (37) and (38). On the other hand, in inflation, $u_k(\eta)$ is expressed as $c_1\, f_k^I(\eta) + c_2\, f_k^{I*}(\eta)$, which is used to calculate $\Phi_I$ and $\xi_I$. From the relations $\Phi_P(\eta_2) = \Phi_I(\eta_2)$ and $\xi_P(\eta_2) = \xi_I(\eta_2)$, the coefficients $c_1$ and $c_2$ can be fixed. Here, the matching condition in which $\Phi$ and $\xi$ ($\Re$) are continuous at the transition point is adopted. However, the matching condition of Deruelle and Mukhanov must be written such that $\Phi$ and $\xi + k^2\Phi/(3(\mathscr{H}^2 - \mathscr{H}'))$ are continuous at the transition point. In the present case of ($k\eta_2 = -z$), the value of $\xi$ becomes smaller than that of the $k^2$ term, causing the latter to dominate at $|z| >> 1$. However, the coefficients $c_1$ and $c_2$ cannot be fixed using $\Phi$ and the $k^2$ term. Thus, the matching condition in which $\Phi$ and $\xi$ ($\Re$) are continuous at the transition point is adopted for the present treatment. As the calculation of the coefficients $c_1$ and $c_2$ is similar to that in the previous sections, only the results are given here.

In the case of the radiation-dominated period before inflation, $|C(k)|$ can be written from equation (12) as

$$C(k) = \frac{\sqrt{\pi}}{43^{3/4}\sqrt{p(p+1)}\sqrt{z}}\, e^{iz/\sqrt{3}p}\, \{(\sqrt{3}+4\sqrt{3}\ p^2 - p(\sqrt{3}+6iz))$$

$$\times\, (H^{(1)}_{-p+1/2}(z)+H^{(2)}_{-p+1/2}(z)) - \sqrt{3}\,(1+p)z(H^{(1)}_{-p+3/2}(z)+H^{(2)}_{-p+3/2}(z))\}. \qquad (39)$$

For $z \to 0$,

$$|C(k)|^2 \cong \frac{2^{-3+2p}\pi(-1+p-4p^2)^2 z^{-2p}}{\sqrt{3}p(1+p)(\Gamma(\frac{3}{2}-p))^2} + \frac{2^{-3+2p}\pi(3-2p-21p^2+24p^3+16p^4)\ z^{-2p+2}}{\sqrt{3}p(1+p)(-3+2p)(\Gamma(\frac{3}{2}-p))^2}. \qquad (40)$$

For $z \to \infty$,

$$|C(k)|^2 \cong \frac{(1+2p+13p^2)+(1+2p-11p^2)\cos\theta}{4\sqrt{3}p(1+p)} + \frac{(-1+9p-3p^2+11p^3)\sin\theta}{4\sqrt{3}(1+p)z}$$

$$+\frac{1+3p-33p^2+61p^3+(1+2p+11p^2-53p^3-4p^4+11p^5)\cos\theta}{8\sqrt{3}(1+p)z^2} \qquad (41)$$

where $\theta = p\pi + 2z$. The quantity $|C(k)|^2$ is plotted as a function of $z$ in figure 4 for $p = -500/499$. $|C(k)|^2$ becomes zero in the $z \to 0$ limit, and oscillates around $(p+1)/(2\sqrt{3}p) \le |C(k)|^2 \le (2\sqrt{3}p)/(p+1)$ in the $z \to \infty$ limit, the latter corresponding to a numerical range of $0.00058 \le |C(k)|^2 \le 1732$ for $p = -500/499$. Comparing figure 1 with figure 4, some differences between the matching conditions are apparent, such as the enhancement of $|C(k)|^2$ from large to small scales, and the marked increase in the amplitude of oscillation as $p \to -1$.

For the case of the scalar matter-dominated period before inflation, $C(k)$ can be written from equation (12) as

$$C(k) = \frac{\sqrt{\pi}}{16\sqrt{3p(p+1)}\sqrt{z^3}}$$

$$i\,e^{2iz/p}\{(p^3 + p^2(-4-2iz)-8iz+p(4+8iz-12z^2))(H^{(1)}_{-p+1/2}(z)+H^{(2)}_{-p+1/2}(z))$$

$$-4(1+p)(p-2iz)z(H^{(1)}_{-p+3/2}(z)+H^{(2)}_{-p+3/2}(z))\}. \tag{42}$$

In the case of $z \to 0$,

$$|C(k)|^2 \cong \frac{2^{-7+2p}\pi p(-2+p)^4 z^{-2-2p}}{3(1+p)(\Gamma(\frac{3}{2}-p))^2} + \frac{2^{-7+2p}\pi(-2+p)^3(24-28p-34p^2+p^3)\ z^{-2p}}{3p(1+p)(-3+2p)(\Gamma(\frac{3}{2}-p))^2} \tag{43}$$

In the case of $z \to \infty$,

$$|C(k)|^2 \cong \frac{(4+8p+13p^2)+(4+8p-5p^2)cos\theta}{12p(1+p)} + \frac{(-4-9p^2+5p^3)sin\theta}{12(1+p)z}$$

$$+\frac{(-2+p)(-4-16p+9p^2+(-4-12p-15p^2+10p^4)cos\theta}{48(1+p)z^2} \tag{44}$$

where $\theta = p\pi + 2z$. $|C(k)|^2$ is plotted as a function of $z$ in figure 5 for $p = -500/499$. $|C(k)|^2$ becomes zero in the $z \to 0$ limit, and oscillates around $2(p+1)/3p \le |C(k)|^2 \le 3p/(2(p+1))$ in the $z \to \infty$ limit, the latter corresponding to a numerical range of $0.0013 \le |C(k)|^2 \le 750$ for $p = -500/499$. $|C(k)|^2$ is large and exhibits oscillation in figure 5, yet maintains a constant value of 1 in figure 3.

The behavior of $|C(k)|^2$ differs considerably between the two matching conditions, and the effect of the factor ($p$ + 1) in the denominator appears to be the main cause. It should also be noted that matching condition B presents a problem in that the case $p = -1$ (de-Sitter inflation) is not applicable for this condition. In de-Sitter inflation, the value of Φ is exactly zero, and in pre-inflation is non-zero. Therefore, a matching value of Φ between pre-inflation and de-Sitter inflation cannot be found. This fact also demonstrates the effect of the factor $p + 1$ of the denominators in equations (39), (41), (42) and (44). This problem only occurs in the case of matching Φ and ξ, and does not occur in the case of matching the gauge potential $u_k(\eta)$.

## 5. Angular power spectrum

Calculation of the power spectra for scalar and tensor perturbations under the two models and matching conditions above for power-law inflation revealed two interesting properties. First, when the length of inflation is finite, the power spectra can be expressed by decreasing functions for $z \to 0$ (in the range of super large scales), and this behavior is seen in all of the cases considered. Second, under matching condition A in the case of a scalar matter-dominated pre-inflation period, the power spectrum becomes 1 for $z \to \infty$, and in the case of radiation-dominated pre-inflation period it oscillates from large to small scales, but the amplitude is constant. Under matching condition B, the power spectra oscillate from large to small scales for both radiation- and scalar matter-dominated pre-inflation, and the amplitude becomes infinite as $p \to -1$ (de-Sitter inflation). These two properties can be observed from figures 1–5, and the details of these properties have been discussed in the preceding papers [11,12].

Here, the effect of this behavior on the angular power spectra is investigated by comparison of the calculated angular power spectra with the WMAP data. The angular power spectrum $C_\ell$ can be written as

$$C_\ell \propto \int d\log k \; P_{\Re}(k) T_\ell(k) T_\ell(k) , \tag{45}$$

where $T_\ell(k)$ is a transfer function. The angular power spectra were computed using a modified CMBFAST code [15] assuming a flat universe and the following parameter values: baryon density $\Omega_b = 0.04$, dark energy density $\Omega_\Lambda = 0.73$, present-day expansion rate $H_0 = 72$ km s$^{-1}$ Mpc$^{-1}$, and reionization optical depth $\tau = 0.17$. The quantities $|C(k)|^2$ and $|C_g(k)|^2$ can be derived analytically using the Bessel function. In the present calculations of the angular power spectra, the expansions of the Bessel function in $|C(k)|^2$ at $z = 0$ and $z = \infty$ were used. The angular power spectra were normalized with respect to 11 data points in the WMAP data from $\ell = 65$ to $\ell = 210$ so as to average the small changes in $\ell$ due to the models and the contribution of oscillation. The value $a$ is adopted as the length of inflation, where $a = 1$ indicates that inflation starts at the time when the perturbations of the current Hubble horizon size exceed the Hubble radius in inflation, that is, the length of inflation is assumed to be close to 60 $e$-folds.

For comparison, the angular power spectrum for the ΛCDM model was also calculated using the above parameters with de-Sitter-like inflation. Moreover, in the case of power-law inflation, the condition $|C(k)|^2 = |C_g(k)|^2 = 1$ is considered with these parameter values. The magnitude of the angular power spectrum for power-law inflation at small $\ell$ ($\ell \leq 100$) changes with $p$, and is notably larger than for the ΛCDM case, that is, the smaller the value of $q$ $(a(\eta) = (-\eta)^p \approx t^q)$, the greater the enhancement of the spectrum. To fit the WMAP data, the value of $p$ must be in the range $-300/299 \leq p < -1$.

The case of radiation-dominated pre-inflation coupled with matching condition A is shown in figure 6 for various values of $a$ at $p = -500/499$, with the ΛCDM model result shown for comparison. Three interesting results can be observed here. The first is that the length of inflation is crucial. Suppression of the cosmic microwave background (CMB) power spectrum is obtained in the case of $a \approx 1$. At $a \leq 1.4$, the angular power spectrum of $\ell = 2,3$ fits the WMAP data within the limit of error. As the decrease occurs smoothly, it seems unlikely that the value of $\ell = 2$ is very small. Secondly, the power spectrum exhibits an oscillation from large to small scales (see figure 1), and the influence of this oscillation on the angular power spectrum is very interesting. In figure 6, the angular power spectra display a small oscillation at $\ell \geq 5$, which may explain the small oscillation seen in the angular power spectrum of the WMAP data. Figure 6 also resolves a small change in the shape of the first peek, indicating a dependence on the value of $a$. Figure 7 shows the cases for $a = 1.0$, 0.9, 0.8, and 0.7 with $p = -500/499$ to illustrate the effect of small changes in $a$. The spectral shape, including the first peak, changes slightly with these small changes in $a$, and such the behavior occurs at $a \approx 50$, whereas the shape of the spectrum remains largely unchanged with changes in $p$. The dependence on $p$ is illustrated in figure 8. To fit the WMAP data, $p$ must be in the range of $-200/199 \leq p \leq -1$, indicating that the contribution of gravitational waves is very small. In the calculation of the angular power spectrum, the scalar-tensor ratio at $\ell = 2$ is needed, which for power-law inflation is written as $R(k) = P_g / P_{\Re} = 16(p+1)/p|C_g(k)|^2/|C\ (k)|^2$. For example, the value of $16(p+1)/p$ is 0.32 at $p = -50/49$.

In the case of matching condition A, $R_c(k) = |C_g(k)|^2/|C\ (k)|^2$ is close to 1 for radiation-dominated pre-inflation and exactly 1 for scalar matter-dominated pre-inflation. However, in the case of matching condition B, $R_c(k)$ has a peculiar shape, approaching zero in the case of radiation-dominated pre-inflation (i.e., $(1+p)/(2\sqrt{3}p)$), but displaying strong peaks at every π value of $z$ with heights of $2\sqrt{3}p/(1+p)$ (see figure 9). The lower limit of $R_c(k)$ appears to be very small, for example, even in the case of $p = -10/9$, the lower limit is $R_c(k) \approx 0.03$. This effect is considered to be accounted for in the integration, since the average value of the integration from $z = 0$ to $z = 10$ is 0.64 at $p = -10/9$ and 0.055 at $p = -500/499$. Thus, in the combination of this effect the scalar-tensor ratio is very small in the case of matching condition B. For example, at $p = -500/499$, $R(k) = (p+1)/p\ R_c(k) \approx 0.032 \times 0.06$.

The results for scalar matter-dominated pre-inflation and matching condition A are shown in figure 10. Similar behavior to that seen here has been derived in many previous papers [6,7]. The length of inflation is again crucial, and suppression of the angular power spectrum is obtained in the case of $a \approx 1$. The angular power spectrum of $\ell = 2,3$ fits the WMAP data within the limit of error at $a \leq 1.2$. As the decrease occurs smoothly, similar to the radiation-dominated case, it is unlikely that the value for $\ell = 2$ is very small. When the length of inflation is longer (see figure 10, $a$ = 10.0), the spectrum becomes similar to that predicted by the ΛCDM model. The power spectrum in the matter-dominated case does not oscillate from large to small scales (see figure 3), and in all 5 scenarios ($a$ = 0.4, 0.8, 1.0, 10.0 and ΛCDM) the spectra become the same shape as that for $\ell \geq 100$ (see figure 10). This demonstrates that the small oscillation of the angular power spectrum in the radiation-dominated case is due to oscillation of the power spectrum. The dependence on $p$ is illustrated in figure 11. To fit the WMAP data, $p$ must be in the range $-200/199 \leq p < -1$, again indicating that the contribution of gravitational waves is very small.

Under matching condition B, similar behavior is obtained for both the radiation-dominated and scalar matter-dominated models. Figure 12 shows the results for $a$ = 1.0 and 10.0 at $p = -500/499$ and for the ΛCDM model for the radiation-dominated case. The suppression of the angular power spectrum at $a \approx 1$ does not occur. Although the power spectra oscillate appreciably from large to small scales under this matching condition (see figures 4 and 5), the oscillatory behavior in figure 12 is not so prominent. However, as a large hump appears in the range $5 \leq \ell \leq 20$ at $a \approx 1$, the angular power spectrum in this case does not fit the WMAP data. At $a \geq 10$, although the angular power spectra do not decrease at $\ell = 2$, the fit with the WMAP data is similar to that achieved by the ΛCDM model. Figure 13 shows the dependence of the spectra on $p$ at $a = 10$. At small $\ell$, the angular power spectra are of higher magnitude than for the ΛCDM model. Thus, in order to fit the WMAP data, $p$ must be in the range $-200/199 \leq p < -1$.

**6. Discussion and summary**

The effect of the length of inflation was investigated by analyzing two pre-inflation models in which pre-inflation is either radiation-dominated or scalar matter-dominated. The effect of pre-inflation was described by the factor $|C(k)|$ (or $|C_g(k)|$), allowing the familiar formulation of the derived power spectrum of curvature perturbations or gravitational waves to be simply multiplied by this factor. These factors were calculated considering two matching conditions, and the characteristics of each of the scenarios were discussed. The corresponding angular power spectra were also calculated for various lengths of inflation ($a$) and values of $p$ and the results compared with the WMAP data.

The power spectrum was found to decrease on super-large scales in all cases. In the case of matching condition A (the gauge potential and its first $\eta$ -derivative are continuous at the transition point), the angular power spectra decreases at $\ell = 2,3$ near $a \approx 1$ , and at $a \leq 1.2$ , the angular power spectra fits the WMAP data within the limit of error (see figures 6,7, and 10). It appears difficult to obtain a good fit of the observed data using Deruelle and Mukhanov's matching condition (matching condition B, see figures 12 and 13).

Under matching condtion A, in the radiation-dominated case the power spectrum of curvature perturbations oscillates from large to small scales. It was shown that this oscillation gives rise to weak oscillation of the angular power spectrum and the shape (including the first peak) of the angular power spectrum changes slightly with the length of inflation (see figures 6 and 7), but not with changes in $p$. Even at large values of $a$ (for example $a \approx 50$), these properties are retained. If the WMAP data exhibits oscillation, this pre-inflation model may be appropriate. Under matching condition B, the oscillation of $|C(k)|$ is very large (see figures 4 and 5 ), yet does not give rise to strong oscillation of the angular power spectrum (see figures 12 and 13).

To fit the WMAP data, both pre-inflation the models require $p$ to be in the range $-200/199 \leq p < -1$ ($q \geq 200$) under matching condition A (see figures 8 and 11). It was shown that this restriction on the value of $p$ is due to the contribution of gravitational waves, which must be very small, suggesting de-Sitter-like inflation. Using the same parameters, the ΛCDM model with power-law inflation yield similar behavior in the caseof $|C(k)|^2$ = $|C_g(k)|^2$=1, that is, the angular power spectra for small $\ell$ ($\ell \leq 100$) is of larger magnitude that for the de-Sitter inflation scenario. Therefore, $p$ values of $-300/299 \leq p < -1$ are required in order to fit the WAMP data. A similar result was derived in Ref. [23] using different values of $p$.

Under matching condition B, however, the angular power spectra do not decrease at $\ell = 2,3$, even if $a \approx 1$, and a large hump occurs at $5 \leq \ell \leq 20$ in this range of $a$, making it impossible to fit to the WMAP data. Although the angular power spectra do not decrease at $\ell = 2,3$ in either inflation model for $a > 10$ and $-200/199 \leq p < -1$, the fit to the WMAP data is similar that achieved by the ΛCDM model, with some oscillation (see figures 12 and 13).

These results indicated that a better fit to the WMAP data is achieved by models with radiation-dominated or scalar matter-dominated pre-inflation, lengths of inflation of $a \le 1.2$, and power-law inflation of $-200/199 \le p < -1$ under matching condition A. It should be noted, however, that the present models do not have a concrete physical base. Nevertheless, it appears that the method employed here is applicable to any physical inflation model with finite length, and similar results are likely to be derived. As an example of a concrete physical model, our group is currently calculating the angular power spectrum for an inflation model described in terms of supergravity, incorporating target-space duality and nonperturbative gaugino condensation in the hidden sector [24], although this example differs from the scenarios considered in the present paper (i.e., the effect of pre-inflation is not considered). A similar analysis is also being investigated for the slow-roll inflation model with finite length.

**Acknowledgments**

The authors would like to thank the staff of Osaka Electro-Communication University for valuable discussions.

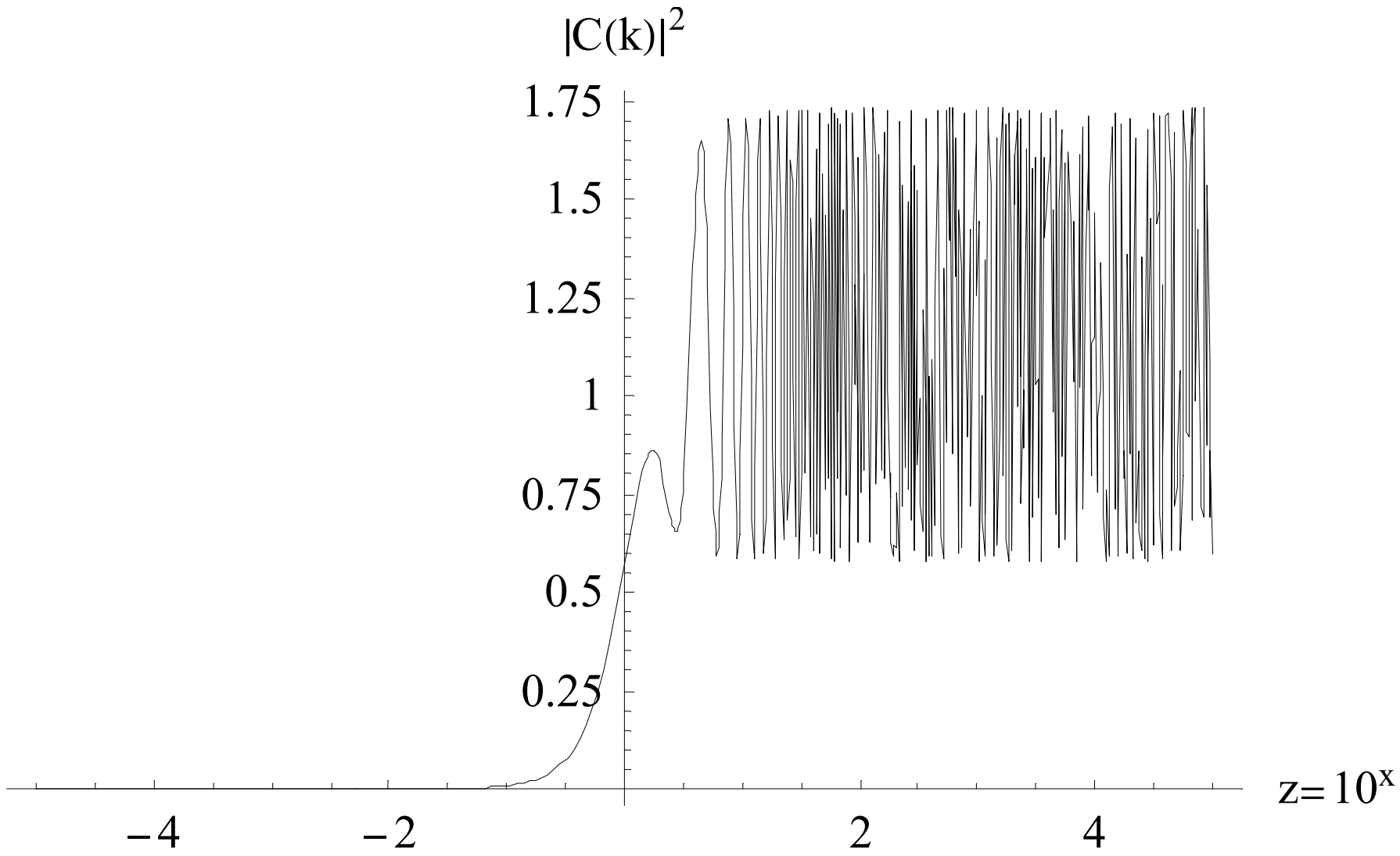

Figure 1. Factor $|C(k)|^2$ as a function of $z$ $(= -k\eta_2)$ for $10^{-5} \le z \le 10^5$ in the case of a radiation-dominated period before inflation under matching condition A ($p$ = –500/499)

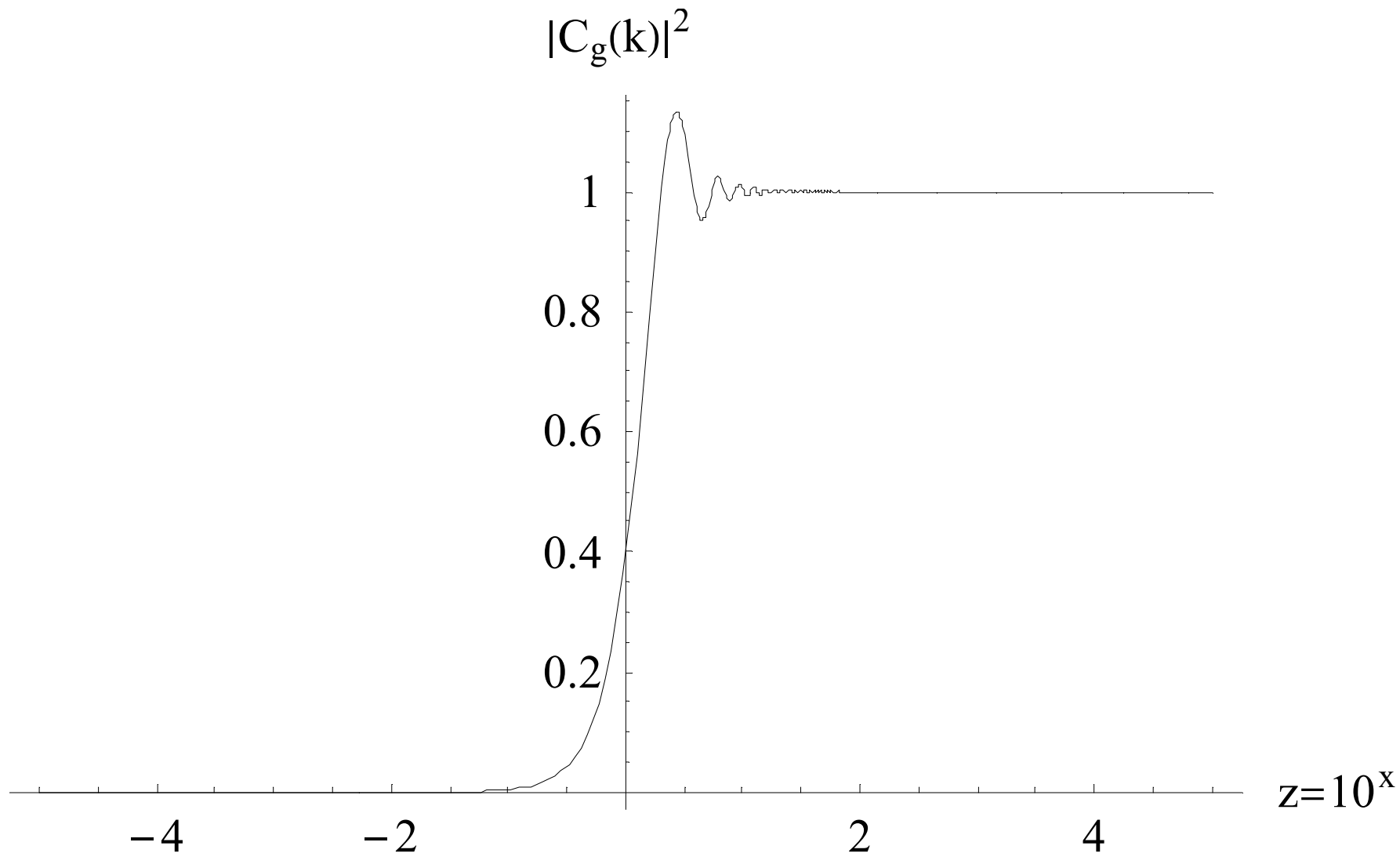

Figure 2. Factor $|C_g(k)|^2$ as a function of $z$ $(= -k\eta_2)$ for $10^{-5} \le z \le 10^5$ in the case of a radiation-dominated period before inflation under matching condition A ($p$ = –500/499)

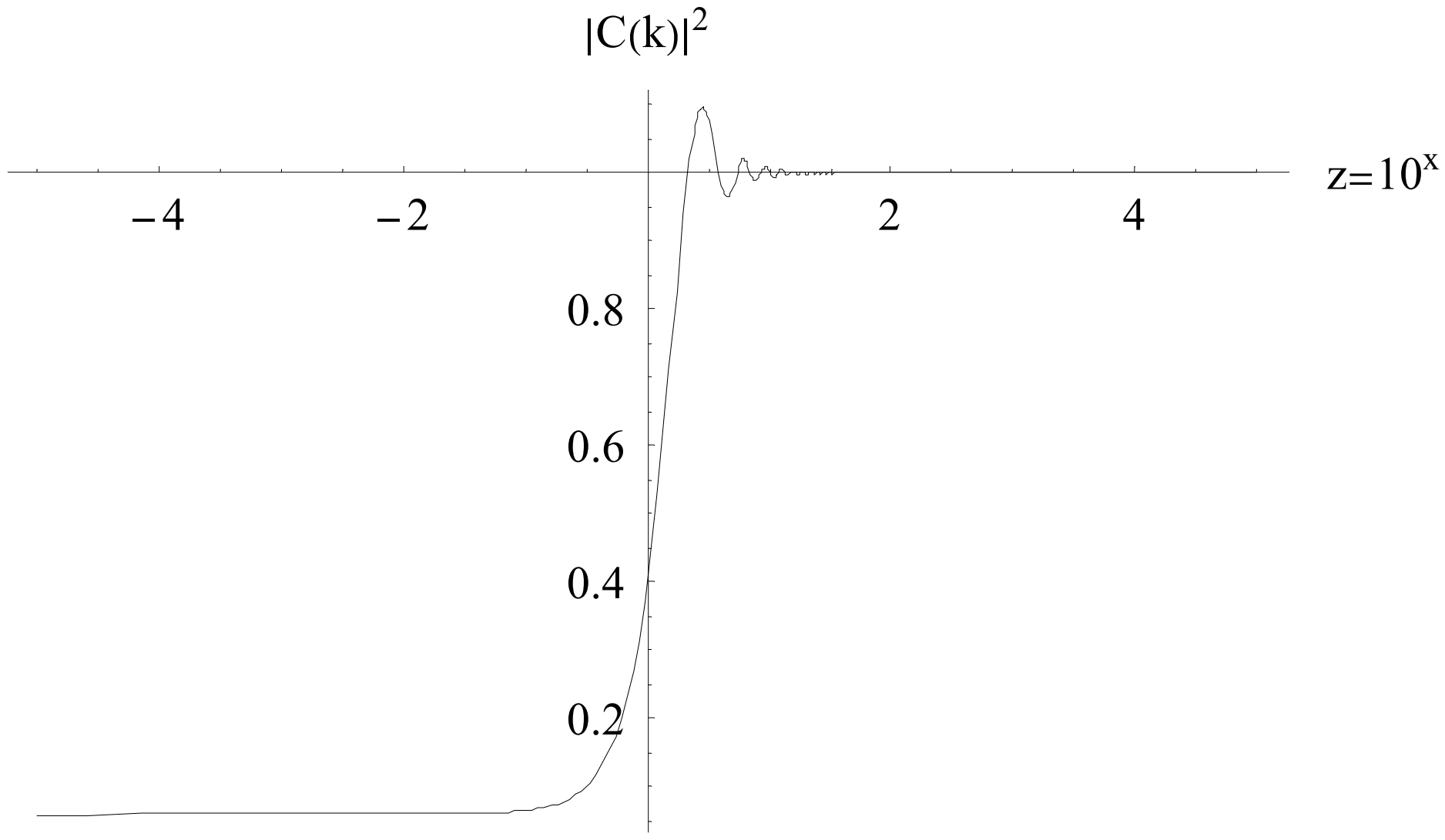

Figure 3. Factor $|C(k)|^2 = |C_g(k)|^2$ as a function of $z$ $(= -k\eta_2)$ for $10^{-5} \le z \le 10^5$ in the case of a scalar matter-dominated period before inflation under matching condition A ($p$ = –500/499)

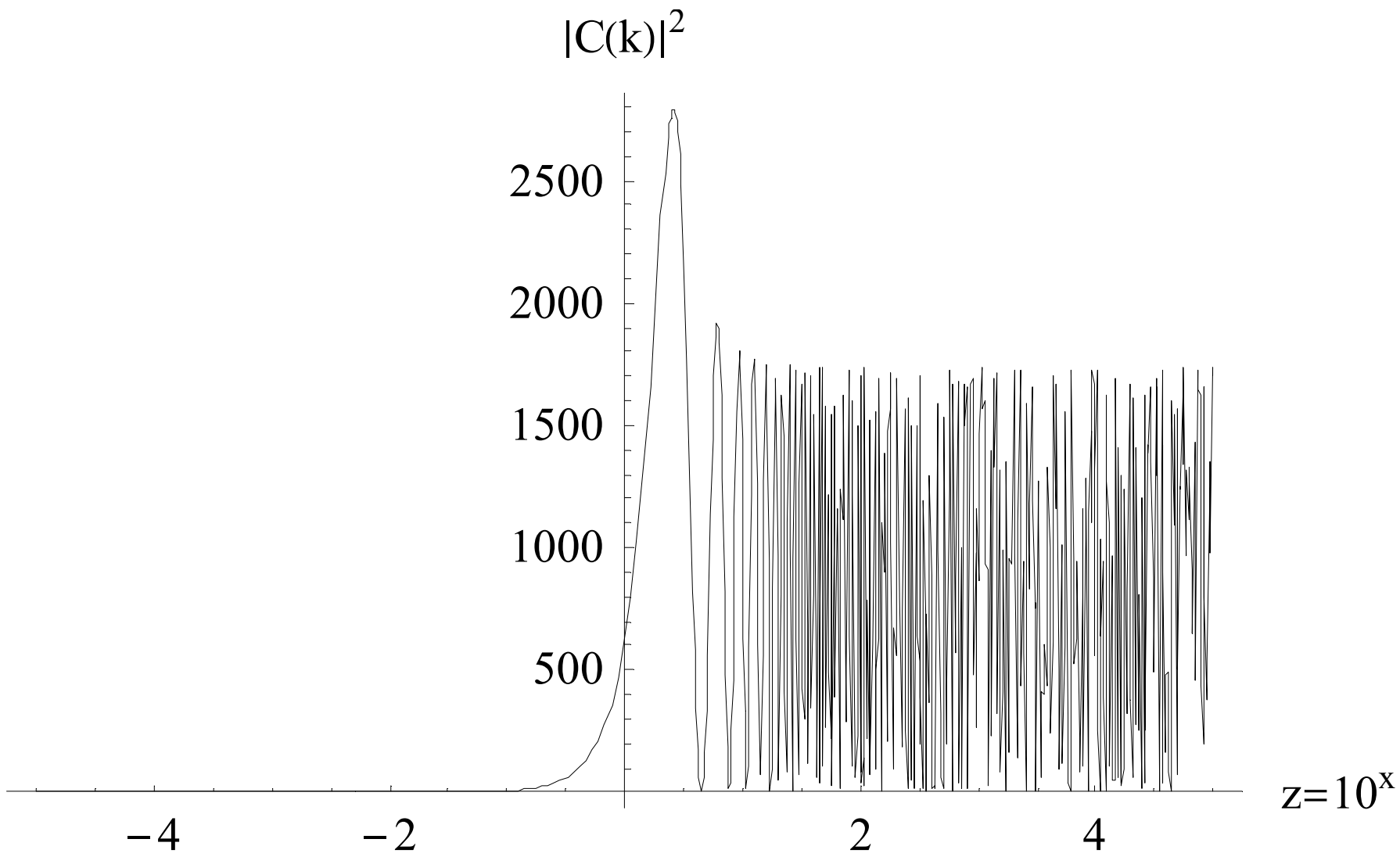

Figure 4. Factor $|C(k)|^2$ as a function of $z$ $(= -k\eta_2)$ for $10^{-5} \le z \le 10^5$ in the case of a radiation-dominated period before inflation under matching condition B ($p$ = –500/499)

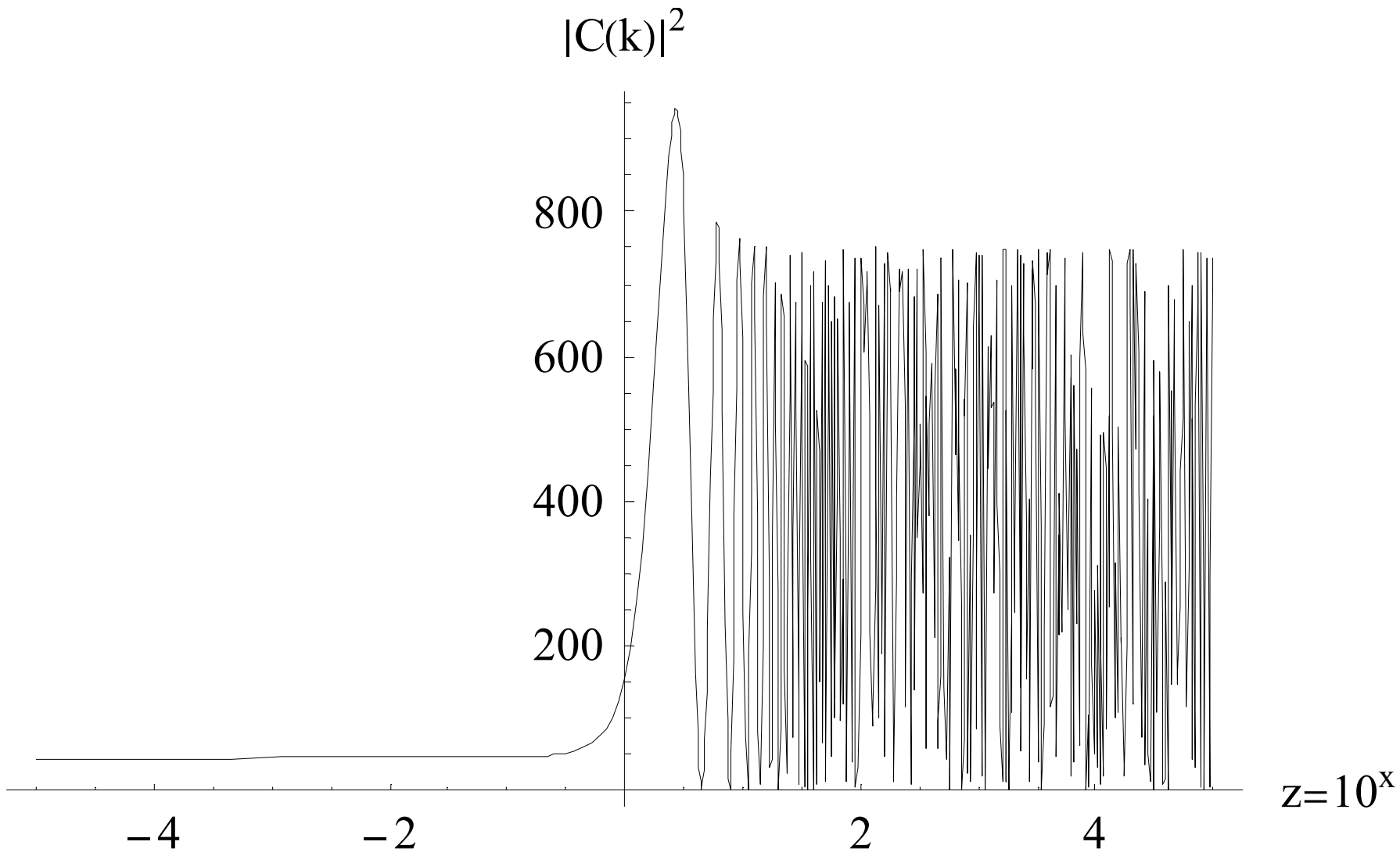

Figure 5. Factor $|C(k)|^2$ as a function of $z$ $(= -k\eta_2)$ for $10^{-5} \le z \le 10^5$ in the case of a scalar matter-dominated period before inflation under matching condition B ($p$ = –500/499)

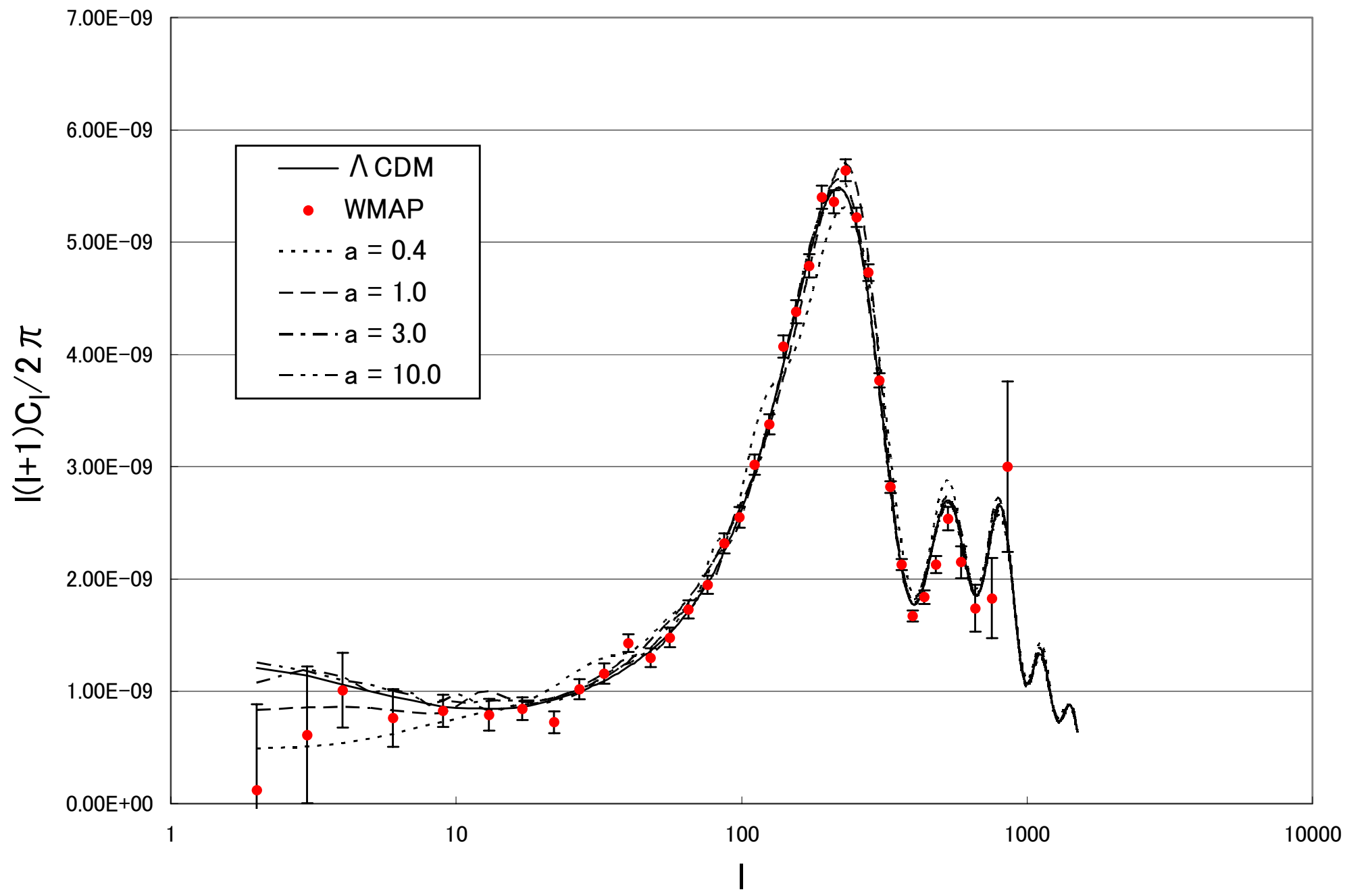

Figure 6. Angular power spectrum in the case of a radiation-dominated period before inflation under matching condition A for $a$ = 0.4, 1.0, 3.0 and 10.0 ($p$ = –500/499) with the ΛCDM model shown for comparison

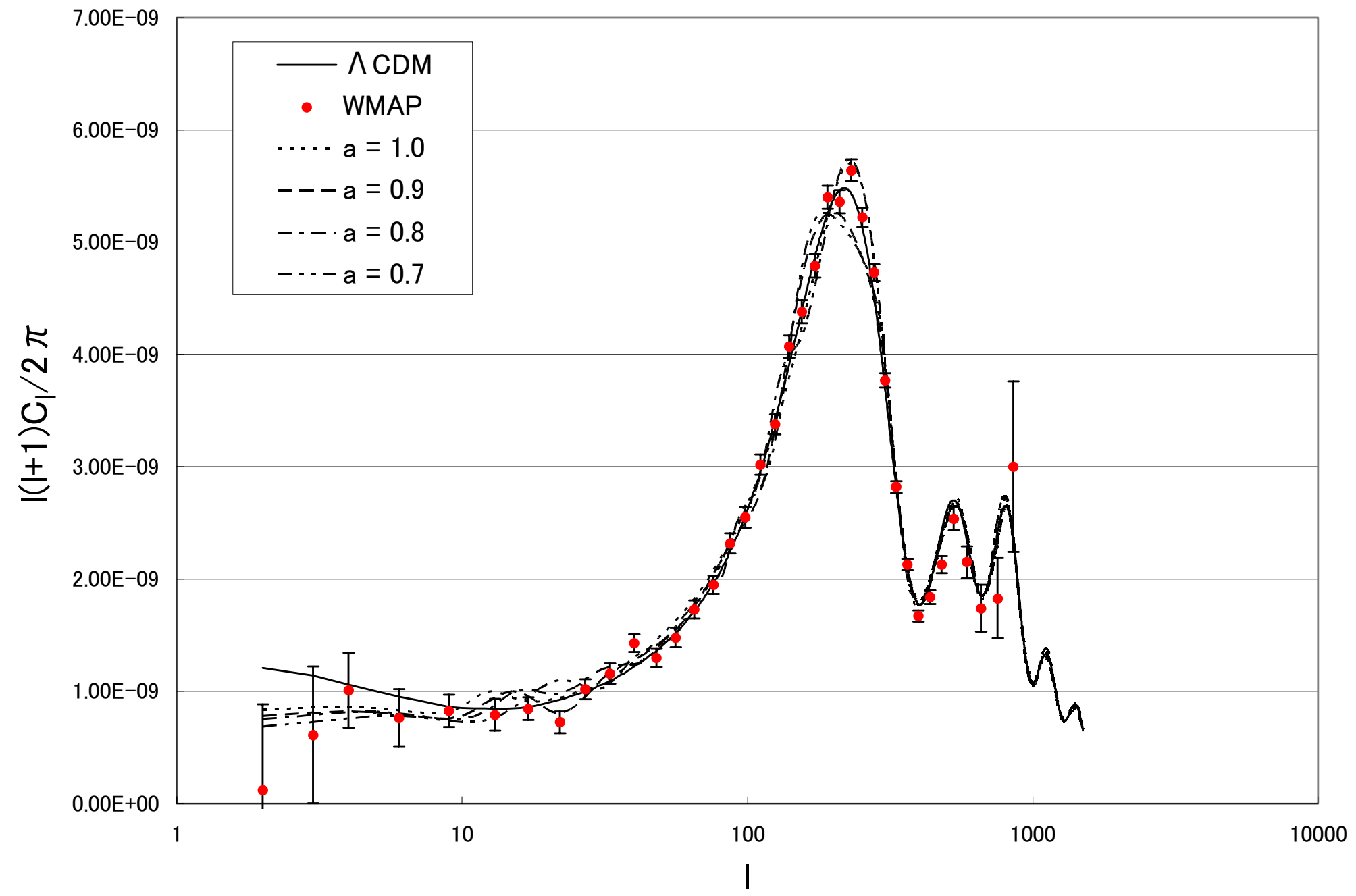

Figure 7. Angular power spectrum in the case of a radiation-dominated period before inflation under matching condition A for $a$ = 1.0, 0.9, 0.8 and 0.7 ($p$ = –500/499) with the ΛCDM model shown for comparison

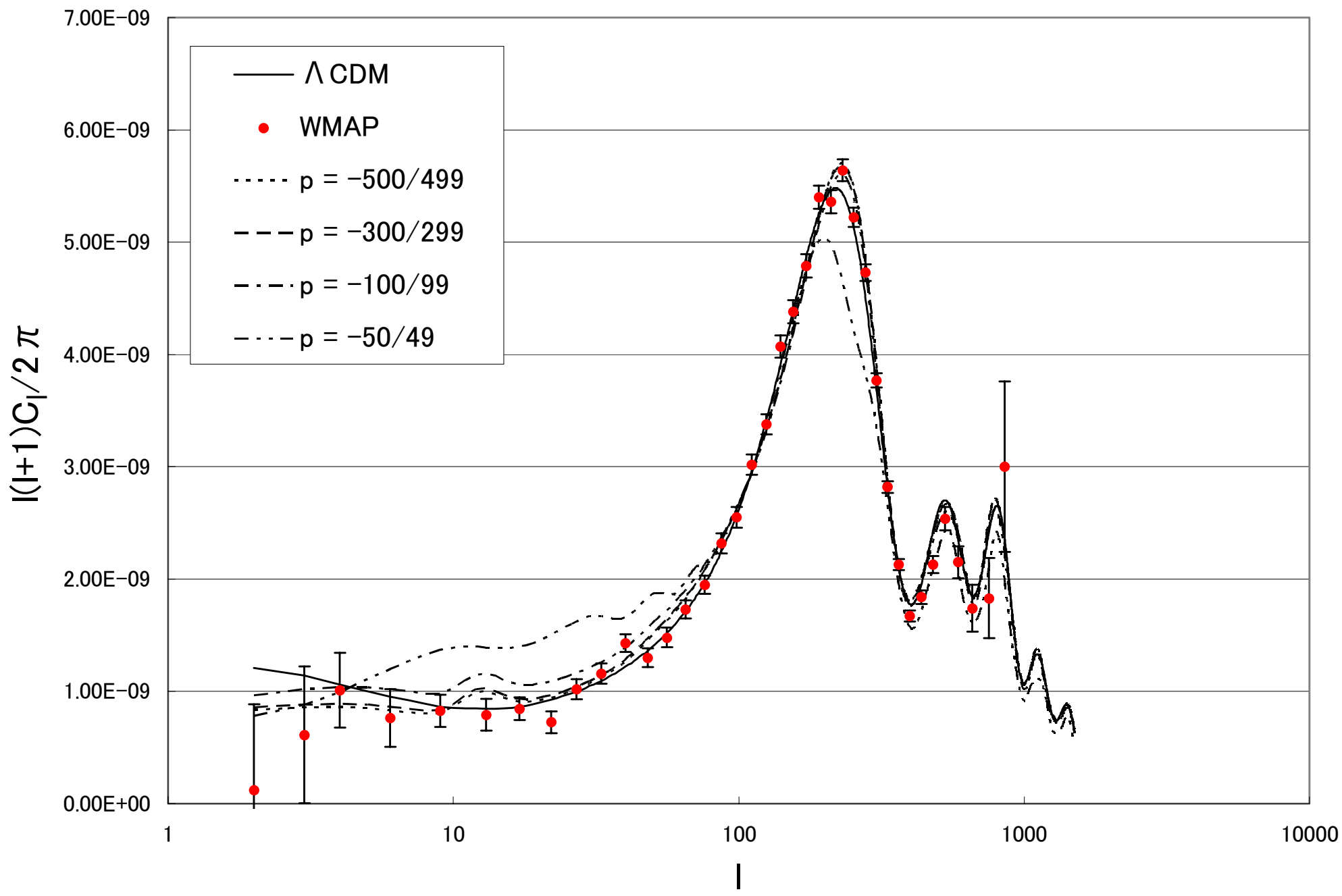

Figure 8. Angular power spectrum in the case of a radiation-dominated period before inflation under matching condition A for $p = -500/499$, $-300/299$, $-100/99$ and $-50/49$ ($a = 1.0$) with the ΛCDM model shown for comparison

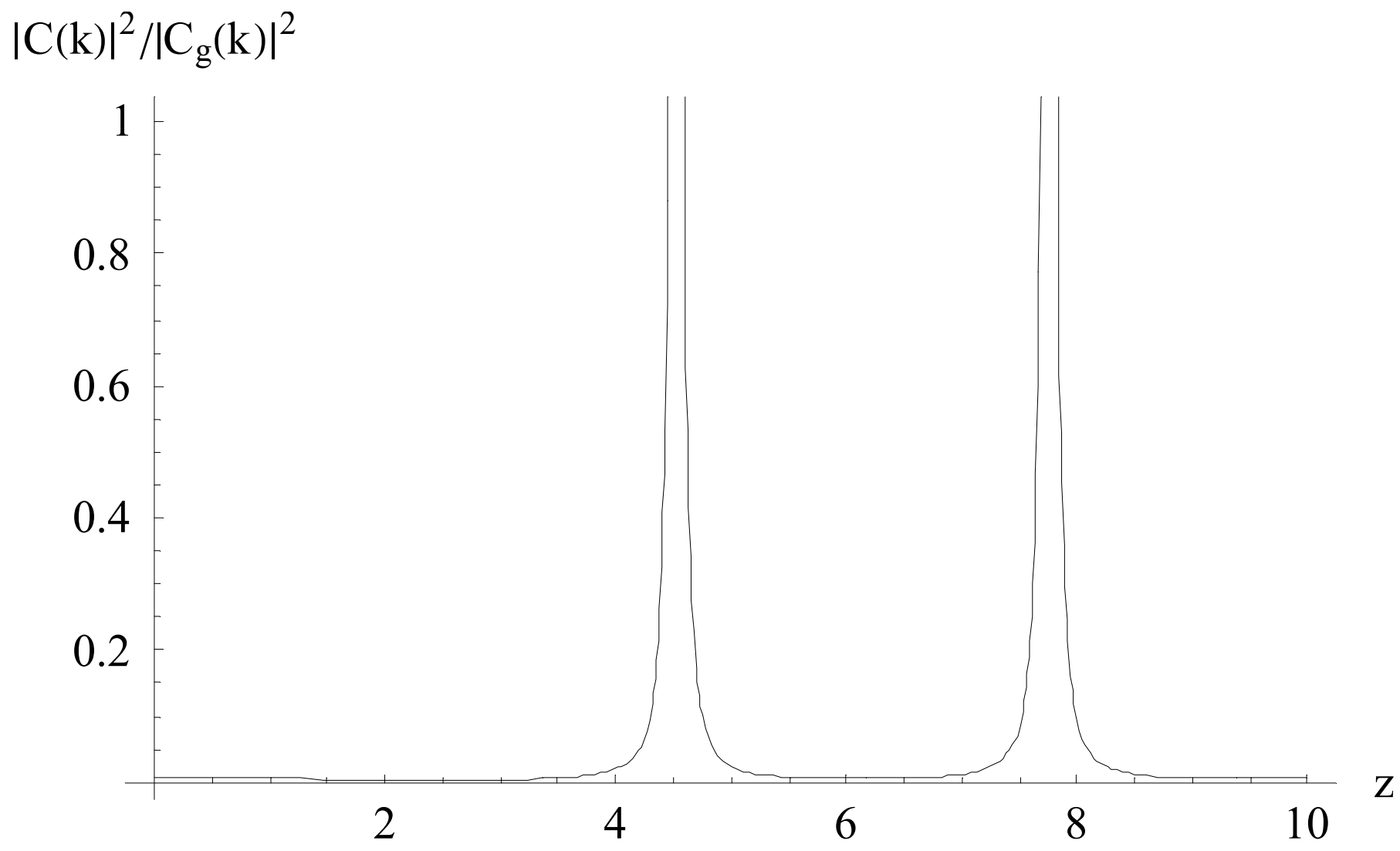

Figure 9. Factor $|C_g(k)|^2/|C(k)|^2$ as a function of $z\ (= -k\eta_2)$ for $0 \le z \le 10$ in the case of a radiation-dominated period before inflation under matching condition B ($p = -50/49$)

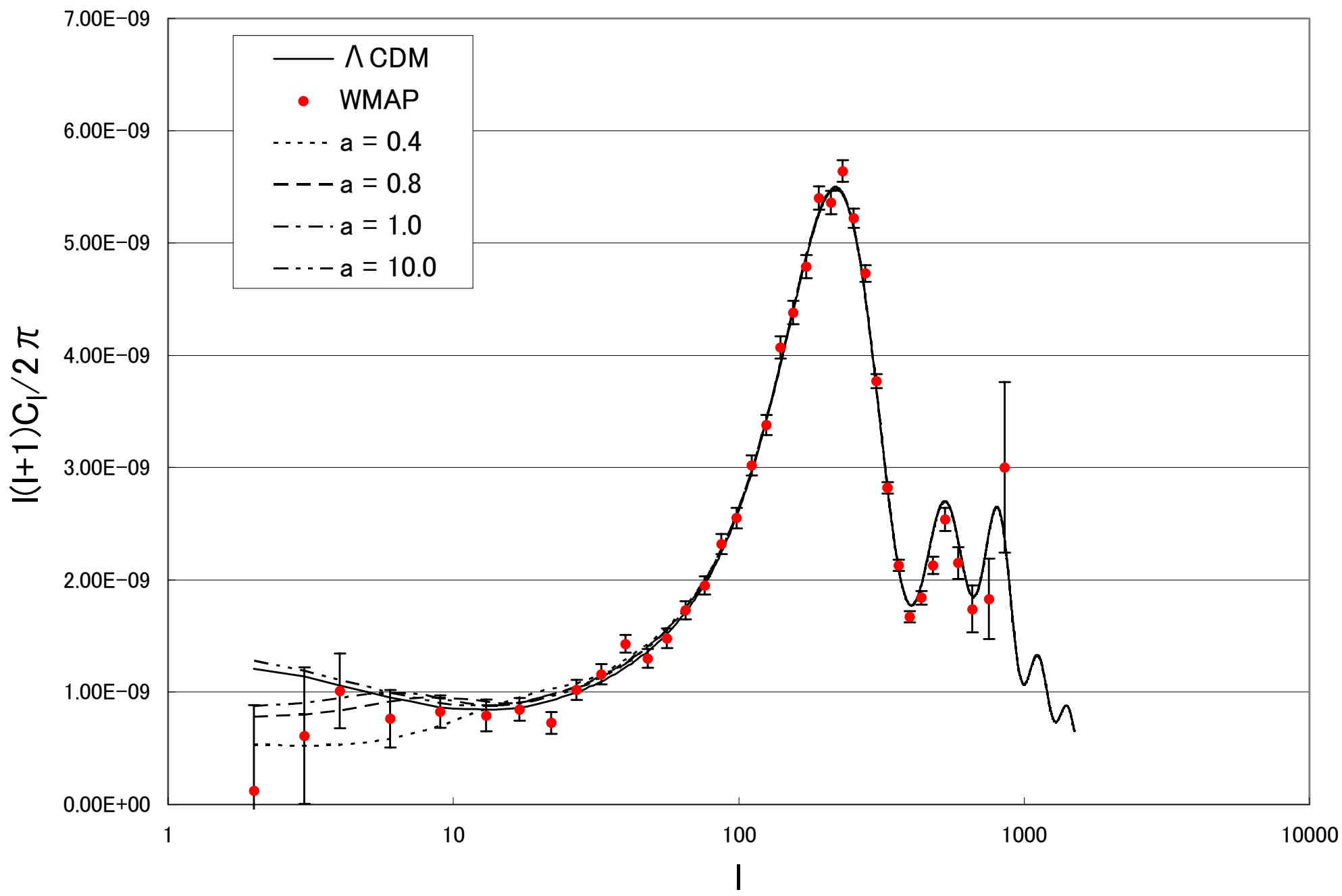

Figure 10. Angular power spectrum in the case of a scalar matter-dominated period before

inflation under matching condition A for $a$ = 0.4, 0.8, 1.0 and 10.0 ($p$ = –500/499) with the ΛCDM model shown for comparison

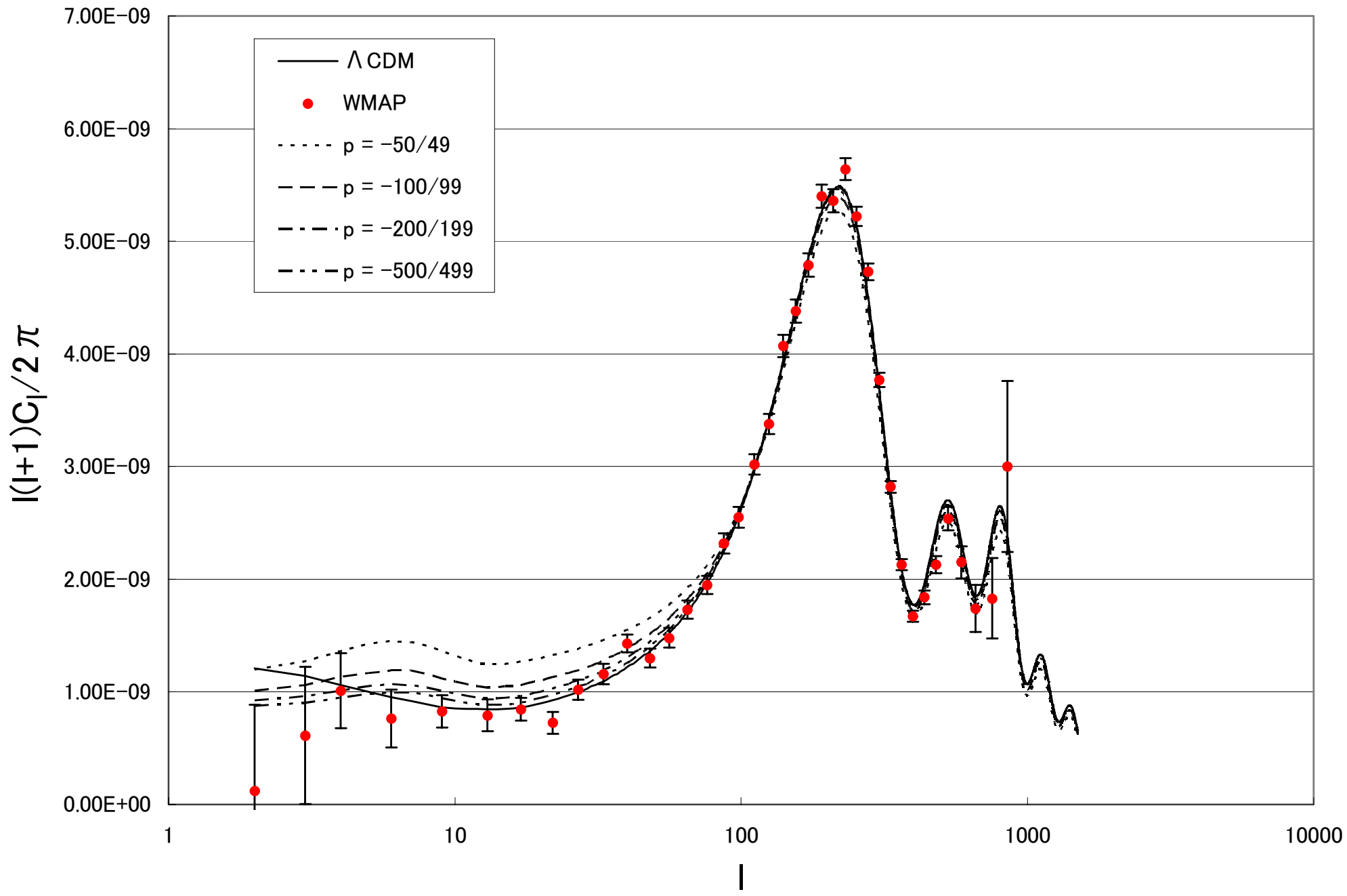

Figure 11. Angular power spectrum in the case of a scalar matter-dominated period before inflation under matching condition A for $p$ = –500/499, –200/199, –100/99 and –50/49 ($a$ =1.0) with the ΛCDM model shown for comparison

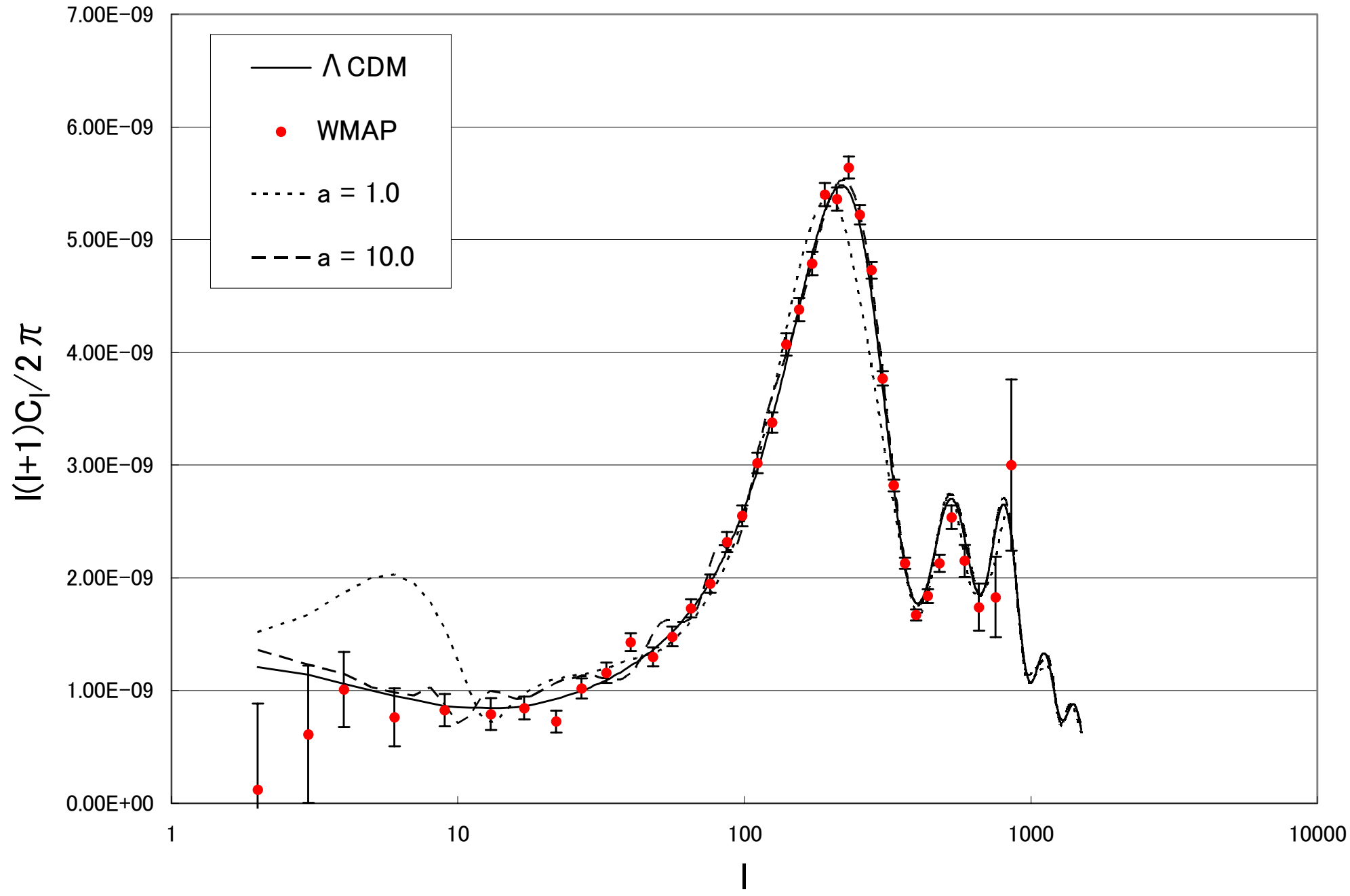

Figure 12. Angular power spectrum in the case of a radiation-dominated period before inflation under matching condition B for $a$ = 1.0 and 10.0 ($p$ = –500/499) with the ΛCDM model shown for comparison

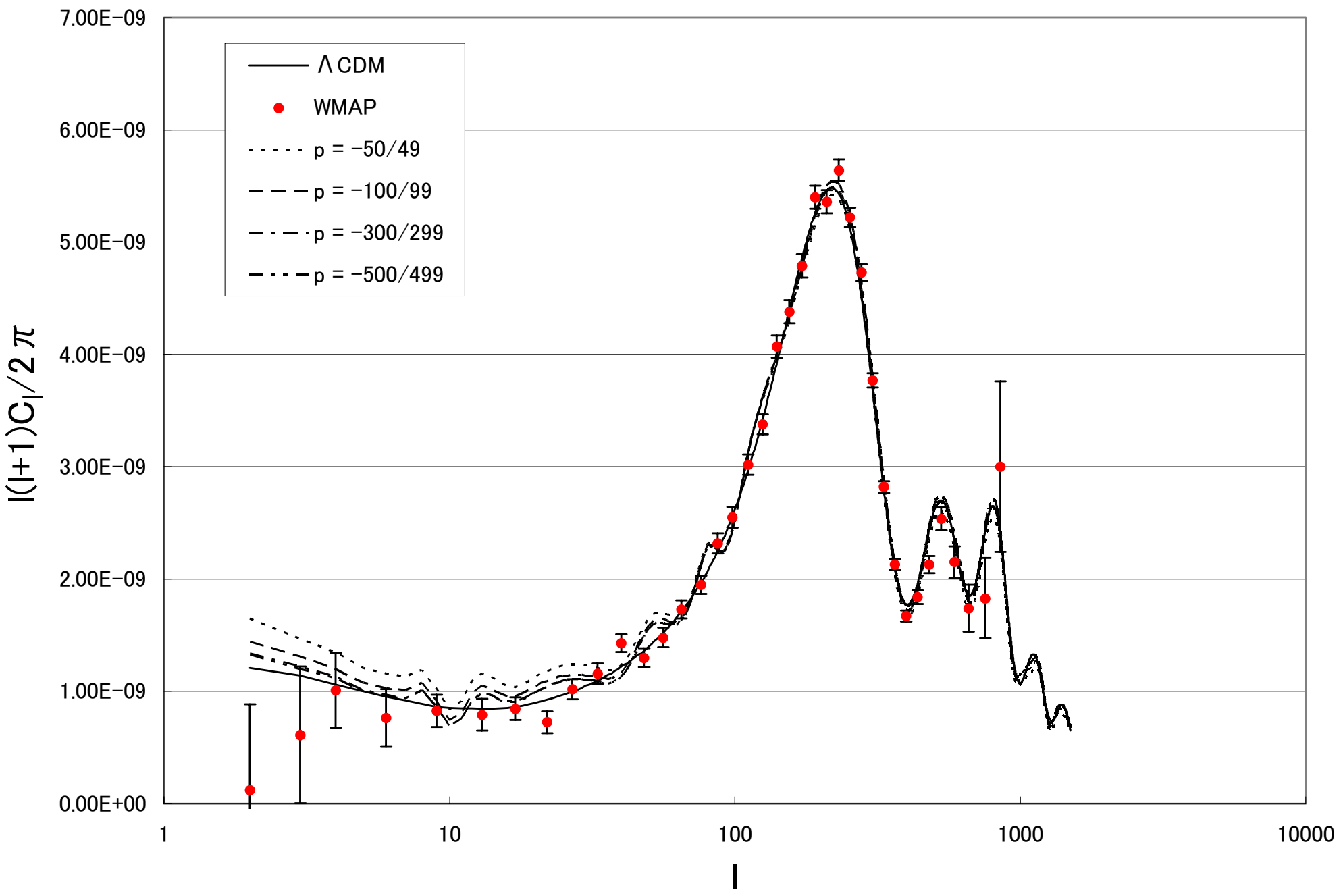

Figure 13. Angular power spectrum in the case of a radiation-dominated period before inflation under matching condition B for $p$ = –500/499, –300/299, –100/99 and –50/49 ($a$ = 10.0) with the ΛCDM model shown for comparison